\documentclass[fleqn,usenatbib]{mnras}
\usepackage{newtxtext,newtxmath}
\usepackage[T1]{fontenc}
\usepackage{ae,aecompl}

\usepackage{amsmath}	
\usepackage{amssymb}	
\usepackage{graphicx,subfigure}

\def\aj{AJ}%
\def\araa{ARA\&A}%
\def\apj{ApJ}%
\def\apjl{ApJL}%
\def\apjs{ApJS}%
\def\ao{Appl.~Opt.}%
\def\apss{Ap\&SS}%
\def\aap{A\&A}%
\def\aapr{A\&A~Rev.}%
\def\mnras{MNRAS}%
\def\nar{New A Rev.}%
\def\pasp{PASP}%
\def\ssr{Space~Sci.~Rev.}%
\def\sa{Science Advances}%
\def\nat{Nature}%
\title[ Fourier time lags in the dwarf nova SS Cygni]{Fourier time lags in the dwarf nova SS Cygni}
\author[E. Aranzana et al.]{E. Aranzana$^{1}$\thanks{E-mail:
E. Aranzana@astro.ru.nl}, S. Scaringi$^{2,3}$, E. K\"{o}rding$^{1}$, V.S. Dhillon$^{4,5}$, D. L. Coppejans$^{6}$\\
$^{1}$Department of Astrophysics/IMAPP, Radboud University Nijmegen, P.O. Box 9010, 6500 GL Nijmegen, The Netherlands\\
$^{2}$School of Physical and Chemical Sciences, University of Canterbury, Private Bag 4800, Christchurch, New Zealand\\
$^{3}$Department of Physics and Astronomy, Texas Tech University, Box 40151, Lubbock, TX 79409-1051, USA\\
$^{4}$Department of Physics and Astronomy, University of Sheffield, Sheffield, S3 7RH, UK\\
$^{5}$Instituto de Astrof\'isica de Canarias, V\'ia Lactea s/n, La Laguna, E-38205 Tenerife, Spain\\
$^{6}$Center for Interdisciplinary Exploration and Research in Astrophysics (CIERA), Northwestern University, Evanston, IL 60208\\
}
\hypersetup{draft}
\begin{document}
\date{Accepted 2018 August 20. Received 2018 August 20; in original form 2017 December 13.}

\pagerange{\pageref{firstpage}--\pageref{lastpage}} \pubyear{2017}

\maketitle

\label{firstpage}
\begin{abstract}
To understand the physical processes governing accretion discs we can study active galactic nuclei (AGN), X-ray binary systems (XRBs) and cataclysmic variables (CVs). It has been shown that XRBs and CVs show similar observational properties such as recurrent outbursts and aperiodic variability. The latter has been extensively studied for XRBs, but only recently have direct phenomenological analogies been found between XRBs and CVs, including the discovery of the rms--flux relation and the optical detection of Fourier-dependent time-lags. We present a Fourier analysis of the well-known CV SS Cyg in quiescence based on data collected at the 4.2--m William Herschel Telescope using ULTRACAM. Light curves in SDSS filters $u'$, $g'$ and $r'$ were taken simultaneously with sub-second cadence. The high cadence and sensitivity of the camera allow us to study the broad-band noise component of the source in the time range $\approx 10000-0.5$ s ($\approx 10^{-4}-2$ Hz). Soft/negative lags with an amplitude $\approx 5$ s at a time-scale of $\approx250$ s were observed, indicating that the emission in the redder bands lags the emission in the bluer bands. This effect could be explained by thermal reprocessing of hard photons in the innermost region of the accretion disc, assuming a high viscosity parameter $\alpha>0.3$, and high irradiation of the disc. Alternatively, it could be associated with the recombination time-scale on the upper layer of the accretions disc.\\
\end{abstract}

\begin{keywords}
accretion discs -- binaries: close -- stars: individual: SS Cyg -- cataclysmic variables
\end{keywords}




\section{Introduction}

Dwarf novae (DNe) are a subclass of cataclysmic variables (CVs) that consist of a primary white dwarf accreting material from a secondary low-mass main-sequence star. The donor overfills its Roche lobe and transfers material onto the compact object through the Lagrangian point L$_{1}$, where an accretion disc is formed to conserve the angular momentum \citep{Frank2002}. Accretion discs are also a common phenomenon in other systems such as X-ray binaries (XRBs) and Active Galactic Nuclei (AGN). Despite the differences in size, central engine and mass, they all share many observational properties. DNe exhibit quasi-periodic outbursts increasing their magnitude up to $\Delta\,\rm m = 2 - 5$ with a duration of $2-20$ days and a recurrence time varying from days to tens of years \citep{Lasota2001discs,Warner2003,Drakeoutburst2014,Coppejans2016}. Outbursts are also a well-known phenomenon observed in XRBs, lasting from days to months \citep{Lewin1995,RemillardXRBs}. Both XRB and CV outbursts are theoretically explained by a sudden onset of thermal and/or viscous instability in the accretion discs surrounding compact objects \citep{Shakura1973,Lasota2001discs}. In XRBs it is common to observe radio emission associated with an out-flowing jet of plasma \citep[e.g.][]{BelloniJets,Fender2004jets}. It is launched during the transition from quiescence to outburst when the spectral state changes from a hard spectrum, dominated by the emission of the corona to a soft spectrum, dominated by the accretion disc. Similar behaviour has been recently reported in the CV SS Cyg, where radio emission associated with a spectral state transition was detected \citep[e.g][]{Kording2008,Russell2016,Mooley2017}. Consistent behaviour is seen in other radio-observed CVs \citep[e.g][]{Coppejans2015,Coppejans2016}. Moreover, all accreting objects show variability on a wide range of time-scales and in a random aperiodic manner or periodic and quasi-periodic variations, e.g. dwarf nova oscillations and quasi-periodic oscillations \citep[e.g.][]{vanDerKlis1989,WarnerWoudt2003,Scaringi2012b}. In particular, the aperiodic variability observed in all these systems is thought to originate in the accretion disc, created by one or several physical processes. The origin is still unclear, and for that reason it is crucial to perform variability studies that can give us insight into the physics governing the accretion flow, the structure and the evolution of the accretion discs.   \\

In the past decades, there have been many studies concentrating on the variability properties of XRBs, in light of the existence of dedicated X-ray missions with very high time-resolution \citep[e.g.][]{XRBsbible}. In the case of CVs, the emission peaks in the optical, thus high-speed optical cameras are required to observe at fast cadences. Well-sampled and long duration observations are also required to study these objects as the characteristic time-scales of CVs are of the order of seconds to days, longer than in XRBs where the time-scales span from milliseconds to hours. This is explained by the fact that the inner disc radius in an XRB lies at a few gravitational radii from the compact object, whilst in CVs it lies at thousands of gravitational radii. \

The accretion connection between XRBs and CVs is further supported by recent discoveries. First, the detection of the linear rms--flux relation in a handful of CVs \citep[see][]{Scaringi2012a,vandeSande2015}, showing that as the source gets brighter it gets more variable. This observational property has been extensively reported in many XRBs and AGN \citep[e.g.][]{Uttley2001,Uttley2005}. Secondly, the discovery by \citet[][]{ScaringiMaccarone2015} that CVs follow the scaling relation that connects the black hole XRBs and the AGN, suggests that the physics of accretion is the same regardless of the nature of the accreting object (see \citet{McHardy2006} and \citet{Koerding2007variaplane} for a detailed description of the variability plane). The scaling relation links the observed break frequency measured in the power spectrum of the light curve with a characteristic time-scale governed by the mass accretion rate and the inner disc radius of the object. This time-scale can be associated to the viscous time-scale because the variability is thought to originate in the disc due to changes in the local viscosity, as explained by the fluctuation propagation accretion model of \citet{Arevalo2006}. According to this model, fluctuations in the local accretion rate occurring further out in the disc (due to changes in viscosity) propagate inwards coupling multiplicatively with the fluctuations in the inner parts of the disc. As a result the variability in the innermost region of the disc is larger than in the outer parts of the disc. Hence, this model can satisfactorily explain the shape of the power spectra, the rms--flux relation and the hard lags observed in XRBs and AGN \citep{Uttley2011Lags}. The latter is explained by the fact that when the fluctuations propagate inwards on the viscous time-scale, they pass from cooler (softer photons) to hotter regions (harder photons) leading to the emission of soft photons before the hard ones.  \\

The detection of soft/negative Fourier time lags in two CVs MV Lyrae and LU Cam has been reported \citep[e.g.][]{Scaringi2013}. In this case the bluer (harder) photons arrive earlier than the redder (softer) photons. More recently, similar lag behaviour was found in four other CVs using cross-correlation functions (CCFs) \citep[e.g.][]{Bruch2015lags}. In particular SS Cygni seems to show negative/soft time lags of the order of seconds. Here, we explore the time lags using Fourier techniques extensively applied in XRBs. Soft time lags have been also detected in many XRBs and AGN but they arise at higher temporal frequencies than the hard lags. They are thought to represent the light crossing time from a variable continuum source to the disc. In AGN, when the variable source illuminates the disc it leads to a soft photo-ionized reflection spectrum, whilst in XRBs the absorbed photons are reprocessed and re-emitted almost instantly as thermal radiation. The soft time lag is of the order of $10-100$ seconds in AGN, whereas in XRBs the delay is of the order of milliseconds. Nonetheless, according to \citet{Scaringi2013} the soft lags observed in CVs can not be explained by the light travel time-scale, because the reflecting region would lie outside the binary orbit. The authors suggest a different origin associated with the thermal time-scale, so that the soft lags are produced by thermal reprocessing of high-energy photons in the accretion disc, or reverse shocks originated close to the compact object. Supporting a reprocessing scenario, it has been shown that the X-ray binary GRO J1655--40 shows a lag of $10-20$ s between the X-rays and the UV/optical emission. The authors suggested the time-scale of reprocessing to likely be the recombination time-scale \citep[e.g.][]{Hynes1998,obrien2002}. \\

In this work we explore the variability properties of the CV SS Cygni in quiescence, one of the brightest dwarf novae known. The source has a magnitude of $\rm V=12$ in quiescence, it is situated at a distance of 114 $\pm$ 2 pc and has an orbital period of 6.6 hours \citep{Bitner2007,MillerJ2013}. The white dwarf mass is about 0.81 M$_{\sun}$, the secondary late-type star is $\sim0.55$ M$_{\sun}$, and the estimated inclination angle lies in the interval of $45^\circ\,\leq\,\rm i\, \leq\,56^\circ$ \citep{Bitner2007}. The mass accretion rate measured in the optical is ${\rm \dot M}\sim\,3\times\,10^{15}\,\rm g\,\rm s^{-1}$  \citep{Patterson1984,Warner1987a}. It exhibits an outburst every approximately 50 days \citep{Szkody1974} during which it is possible to detect coherent dwarf nova oscillations of $10-30$ s and quasi-periodic oscillations \citep[][]{Pretorius2006}. \\
\begin{table*}
\centering
\caption{Observation log of SS Cyg with ULTRACAM.}
\begin{tabular}{l c c c c c}
		\hline\hline
		Date & Run & Start UTC & End UTC & Exposure/frame (s) & Frames \\
		\hline
	2013 August 1 & 1 & $21:56:40$ & $05:47:06$ &0.2393 & 117933\\
	2013 August 2 & 2 & $00:38:17$ & $02:17:26$ & 0.3215 & 17530\\
	2013 August 2 & 3 & $02:17:28$ & $03:56:32$ & 0.3215 & 17515\\
	2013 August 2 & 4 & $03:56:33$ & $05:35:53$ & 0.3215 & 17563\\
		\hline
		\end{tabular}
		\label{table:1}
		\end{table*}
		
\begin{figure*}
\centering
\includegraphics[width=15cm]{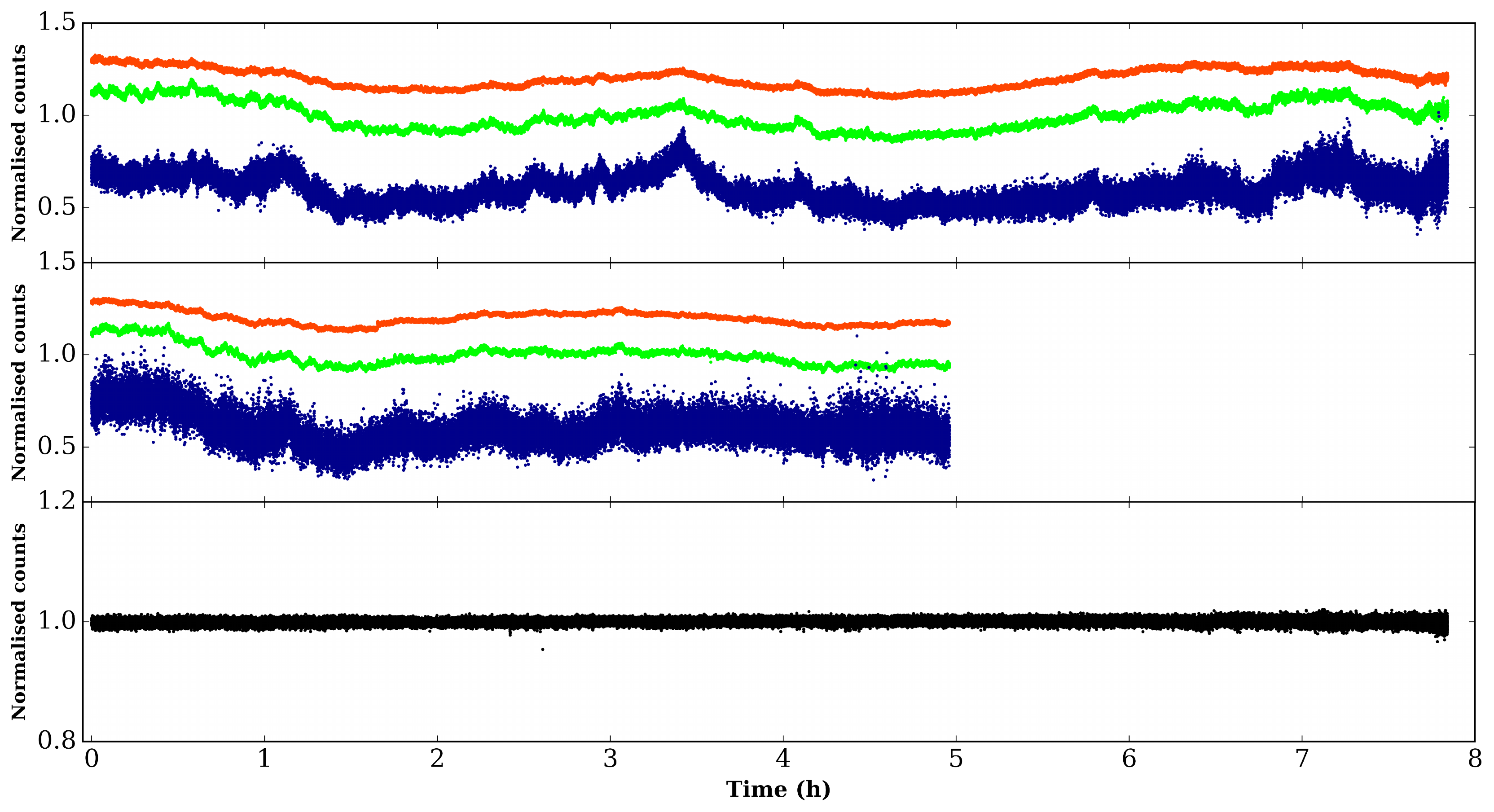}
\caption{The two upper panels show the light curves of SS Cygni observed during two consecutive nights with ULTRACAM in three colours simultaneously: $r'$, $g'$ and $u'$ . The first night was of $\sim 8$ h duration and the second night SS Cyg was observed during $\sim 5$ h divided in three consecutive runs. For comparison with SS Cyg, the bottom panel shows the light curve of one of the comparison stars TYC 3196-857-1 in the $r'$ band the first night, illustrating the lack of variability compared with SS Cygni. }
\label{Fig:lcr}
\end{figure*}
The paper is organised as follows: the description of the observations, the data reduction and the analysis of the light curves are described in Section \ref{sec:obs}. The results are described in Section \ref{sec:results}, followed by a discussion in Section \ref{sec:discussion} and a summary in Section \ref{sec:summary}. \\


\section[]{Observations and data analysis}
\label{sec:obs}

We observed SS Cygni when it was in quiescence with ULTRACAM \citep{Dhillon2007} mounted at the Cassegrain focus of the 4.2--m William Herschel Telescope (WHT) on La Palma. ULTRACAM is a high-speed triple-beam CCD camera designed to image faint astronomical objects at high temporal resolutions. It enables the user to observe a target in three optical filters simultaneously with a cadence of up to 0.0033 s (300 Hz). The instrument consists of three frame-transfer chips e2v 47-20 CCDs of 1024x1024 $\rm pixel^{2}$ area, providing a 5 arcminute field at a scale of 0.3 arcsecond/pixel. Incident light from the focal plane of the telescope is collimated and then split by two dichroic beam-splitters into three beams, defined by the SDSS {\it u'} (3543 \AA), SDSS {\it g'} (4770 \AA) and SDSS {\it r'} (6231 \AA) filters. The dead time between exposures is negligible (25 msec) due to the frame-transfer chips. The read-out speed was set to fast and no binning was applied. The observations took place on 2013 August 1 and 2. In Table \ref{table:1} we show the observing log. The first run is the longest, with a total duration of about 8 h and with the highest cadence of $\sim$ 0.24 s. During the second night the cadence had to be reduced in order to maintain a good signal-to-noise in the poorer conditions and we had to check it in between, as a consequence of this the data is in three different runs all taken with the same cadence. Flat fields were taken with the telescope spiralling of the twilight sky before and after the observations. Six--windowed mode was used for the target, and 5 comparison stars to perform differential photometry.\
        
The data were reduced using the ULTRACAM pipeline reduction package. Every science frame was de-biased and then flat-fielded. We used variable-sized apertures scaled to the seeing for each filter, and we used the normal aperture extraction and applied {\it moffat} profiles. In individual frames of the {\it u'} filter only two comparison stars were bright enough to extract, while for the $r'$ and $g'$ filters we used five. In Fig. \ref{Fig:lcr} we present the light curves for the two nights separately in the upper two panels. We present the normalised counts per cadence of SS Cyg by using the comparison stars to correct for atmospheric variations. An off-set of $\pm0.2$ counts is added for clarity. In the lower panel we show the light curve in the $r'$ band of the comparison star TYC 3196-857-1 with ICRS coordinates $\alpha=21^{\rm h}42^{\rm m} 27.1695^{\rm s} $ and $\delta=+43^{\circ}33'44.601''$ divided by the other comparison stars. Whilst SS Cygni shows strong aperiodic variability in all three colours, the solar-type comparison star is constant.\\

\subsection{Fourier analysis of the light curves}
\label{sec:Fourier}

\begin{figure}
\centering
\includegraphics[width=7.5cm]{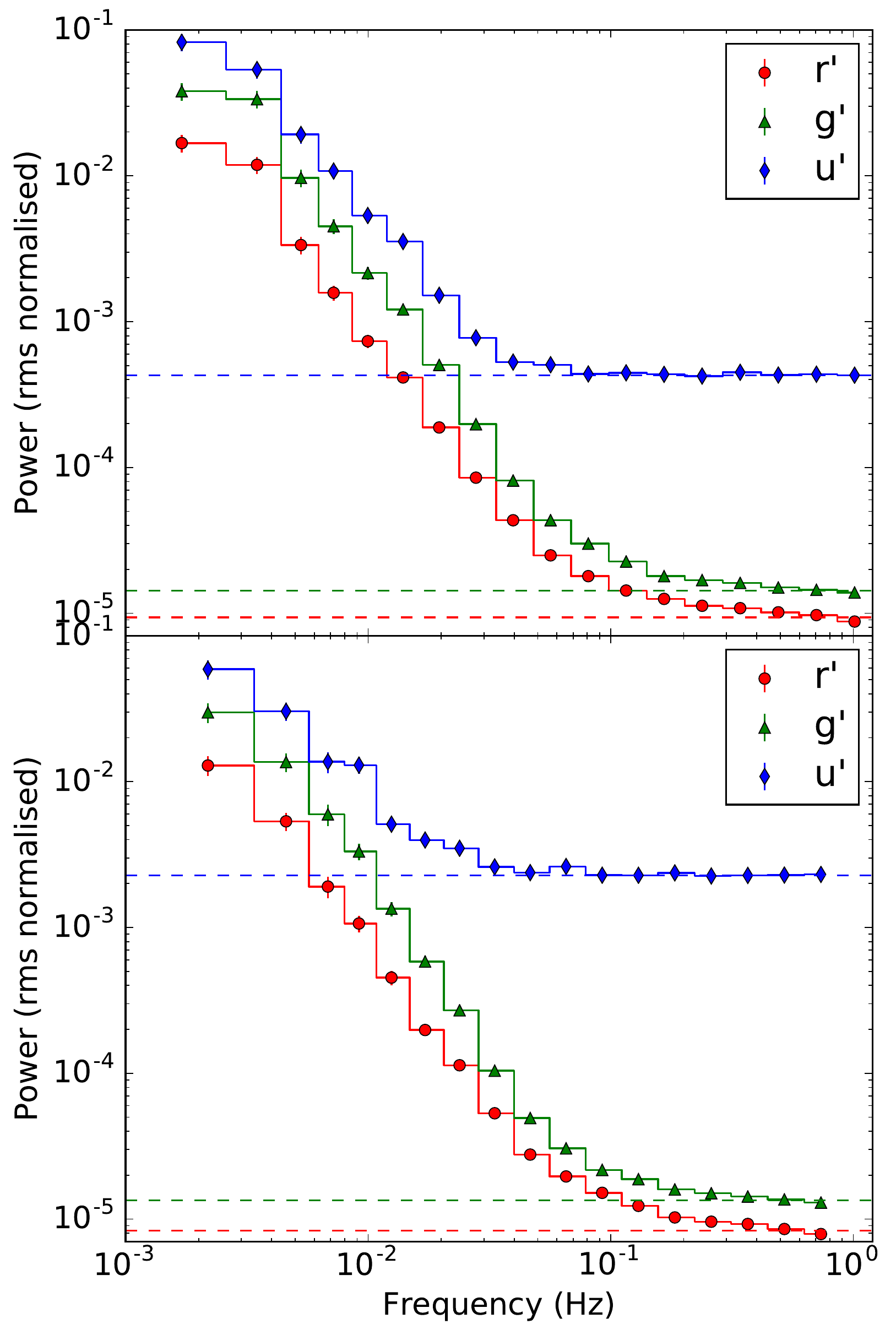}
\caption{Power spectrum of SS Cyg on the 1st August (upper panel) and 2nd of August (lower panel). The blue diamonds denote the $u'$ band, green triangles the $g'$ band and the red circles the $r'$ band. The horizontal dashed lines indicate the noise level for each  colour. The power spectra show a time-scale break at $333$ s ($\sim3\times10^{-3}$ Hz) followed by a steep power-law with power law index $> -2.0$. }
\label{Fig:psd}
\end{figure}

We performed a Fourier analysis of the light curves using timing techniques extensively applied in the X-rays and described in \citet{vanDerKlis1989b} and \citet{Nowak1996}. \citet{Scaringi2013} performed a similar analysis as to the one we describe below when studying the CVs MV Lyr and LU Cam. It is important to mention that the light curves are evenly sampled in each run, thus we used the Fast Fourier Transform (FFT). First, we split the light curve into non-overlapping segments of equal duration and applied the FFT to each individual segment. Here we split each run into segments of $\approx48$ minutes ($\sim12.000$ data points), so that for the first night we used nine segments and for the second night six segments (hereafter $k_{1}$ and $k_{2}$, respectively). It is important to note that we performed the Fourier analysis for the two nights separately because they have different sampling as shown in Table \ref{table:1}. Defining $x_{u,i} (t)$ and $x_{r,i} (t)$ as the light curve segments $i$ in the $u'$ and the $r'$ bands, we computed the FFT to obtain the Fourier amplitudes, $X_{u,i} (f)$ and $X_{r,i} (f)$. The power spectra (PSD) were then calculated as $P_{u,i}(f)=|X_{u,i}|^2$  and $P_{r,i}(f)=|X_{r,i}|^2$, respectively. For each night we obtained the power spectrum of each segment independently and then we averaged them. Later we applied logarithmic binning and derived the PSD errors for each bin. The uncertainty is calculated as the power divided by the square root of the number of measurements used in the specific bin, $ m= k_{1,2}\times N$, where $N$ is the number of data points in each bin and $\rm k_{1,2}$ the number of segments used in the first or the second night. We applied the rms normalisation described in \citet{Belloni1990} to the power spectra, so that the square root of the integrated PSD over a frequency range yields the fractional amplitude of variability (see Fig. \ref{Fig:psd}). \\

Subsequently, we performed a cross spectral analysis for each segment calculated as $C_{i}(f)=X_{r, i}^{*}(f)X_{u,i}(f)$, where $X_{r, i}^{*}$ is the complex conjugate. We averaged all the cross spectra and log-binned the average (for each night separately) obtaining $C(f)$. The phase lag between the two simultaneous light curves, $\phi(f)$, is the argument of the complex-valued cross spectrum $C(f)$. The time lag, $\tau (f)$ is computed by dividing the phase lag $\phi(f)$ by $2\pi \rm f$. Thereafter, we calculated the raw coherence function $\gamma^{2}_{\rm raw}$, which is a measure of the degree of correlation between two simultaneous light curves in different bands:

\begin{equation}
\gamma_{raw}^{2}=\frac{|C(f)|^{2}}{P_{r}(f)P_{u}(f)},
\label{eq:raw}
\end{equation}
\noindent where $P_{r}(f)$ and $P_{u}(f)$ are averaged and log-binned. The statistical uncertainty is $\delta\gamma_{raw}^{2}=(2/m)^{1/2}(1-\gamma_{raw}^{2})/|\gamma_{raw}|$. The uncertainty of the phase lag is calculated as:

\begin{equation}
\delta\phi(f)  = \frac{1}{\sqrt{m}}\sqrt{\frac{1-\gamma_{raw}^{2}}{2\gamma_{raw}^{2}}},
\label{Eq:phaseerror}
\end{equation}

\noindent and the error of the time lag is defined as $\delta\tau = \delta\phi(f)/2\pi f$. However, since the coherence of a real signal will be affected by the noise, every single term in Eq. \ref{eq:raw} needs to be corrected from counting noise. That is, correcting the powers in the denominator of the equation by the measured Poisson noise level and then correcting the cross-spectra. The detailed description to compute the intrinsic coherence $\gamma^{2}$ and its error, $\delta\gamma^{2}$ can be found in Eq. 8 of \citet{Vaughan1997}.\

The description given above was only for the $r'$ and $u'$ combination. We followed the same procedure to derive the time lags and the coherence for the other two colour combinations, $r'$ and $g'$, and $g'$ and $u'$. As mentioned before, the cross spectral analysis was performed for the two nights separately, obtaining the raw and the intrinsic coherence and the time lags. To improve the reliability of the result we combined the two nights and computed the phase/time lags and the coherence. For this, we first created an array containing the unbinned time lag values of the two nights and sorted them in frequency. Later we applied a logarithmic binning with the condition of a minimum number of data points per bin, where $N> 5$. Therefore, each bin contains a  certain number of data points that correspond to the first night $N_{1}$ and to the second night $N_{2}$, so that $N_{1}+ N_{2} > 5$. The total number of data points per bin is calculated as $m = k_{1}\times N_{1}+ k_{2}\times N_{2}$. To combine the phase lags of the two nights we proceeded in the same way. In order to combine the raw and the intrinsic coherence of the two nights we log-binned all the terms of equation \ref{eq:raw} and Eq. 8 of \citet{Vaughan1997}, in the same manner as we described for the time lags.   \\

\begin{figure}
\centering
\includegraphics[width=8cm]{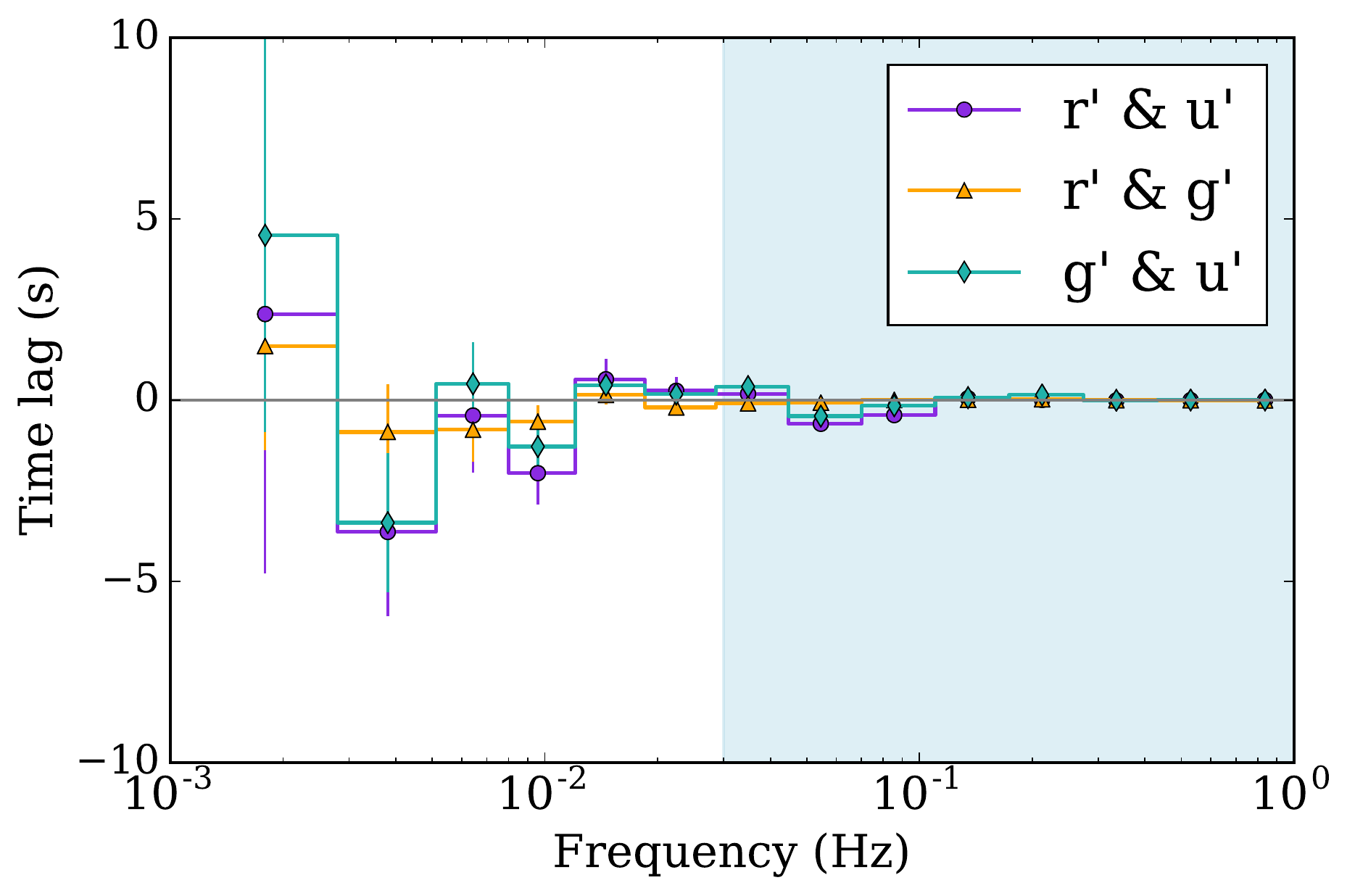}
\includegraphics[width=8cm]{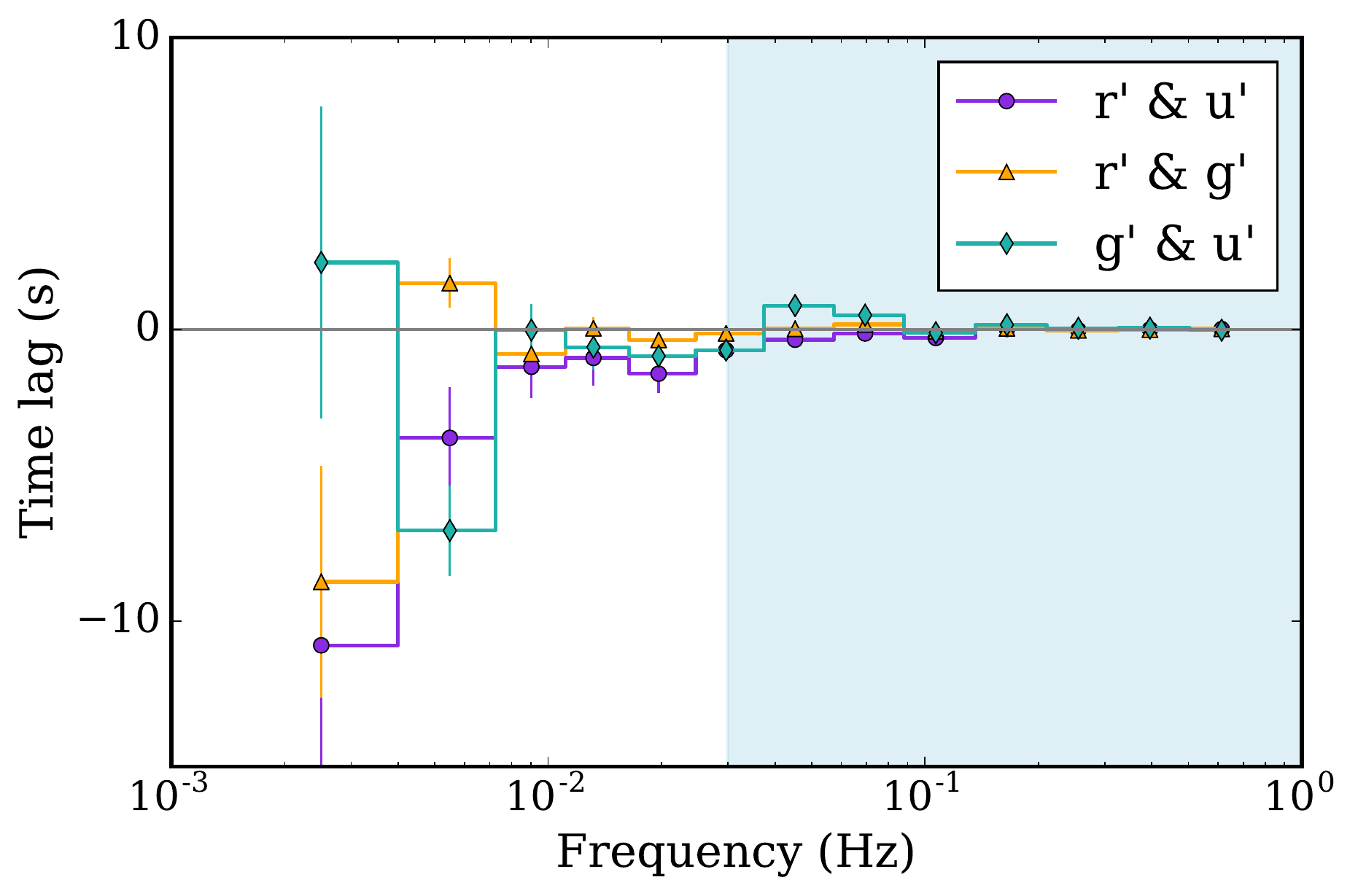}
\caption{{\it Upper panel:} Time lags observed on the first night. A tentative soft/negative lag is observed at $4\times10^{-3}$ Hz indicating that  emission from the blue bands is emitted earlier than the redder band. {\it Lower panel:} Time lags observed during the second night. In both figures the purple circles denote the $r'$ and $u'$ combination, the green diamonds the $g'$ and $u'$ and the orange triangles the $r'$ and $g'$ combination. The shaded region indicates the frequencies that are dominated by Poisson noise.}
\label{Fig:lags1}
\end{figure}


\section{Results}
\label{sec:results}

The PSDs of the two nights is shown in Fig. \ref{Fig:psd} in units of $(\rm rms/mean)^{2}/\rm Hz$. The power in the $u'$ band is higher than in the other bands because the variability in this wavelength is thought to arise from a hotter region, closer to the white dwarf. The horizontal dashed lines indicate the noise level for each colour. The Poisson noise in the $u'$ band clearly dominates above $0.03$ Hz. The $r'$ and the $g'$ binned PSDs that are less dominated by Poisson noise can be fit with a bending power-law with the time-scale break at $333$ s ($\sim 3\times10^{-3}$ Hz) and a high-frequency slope steeper than $-2$. Since the Poisson noise dominates in this source, thus at time-scales $<33$ s ($\nu>0.03$ Hz) we cannot extract any meaningful result. \\

\begin{figure}
\centering
\includegraphics[width=8.0cm]{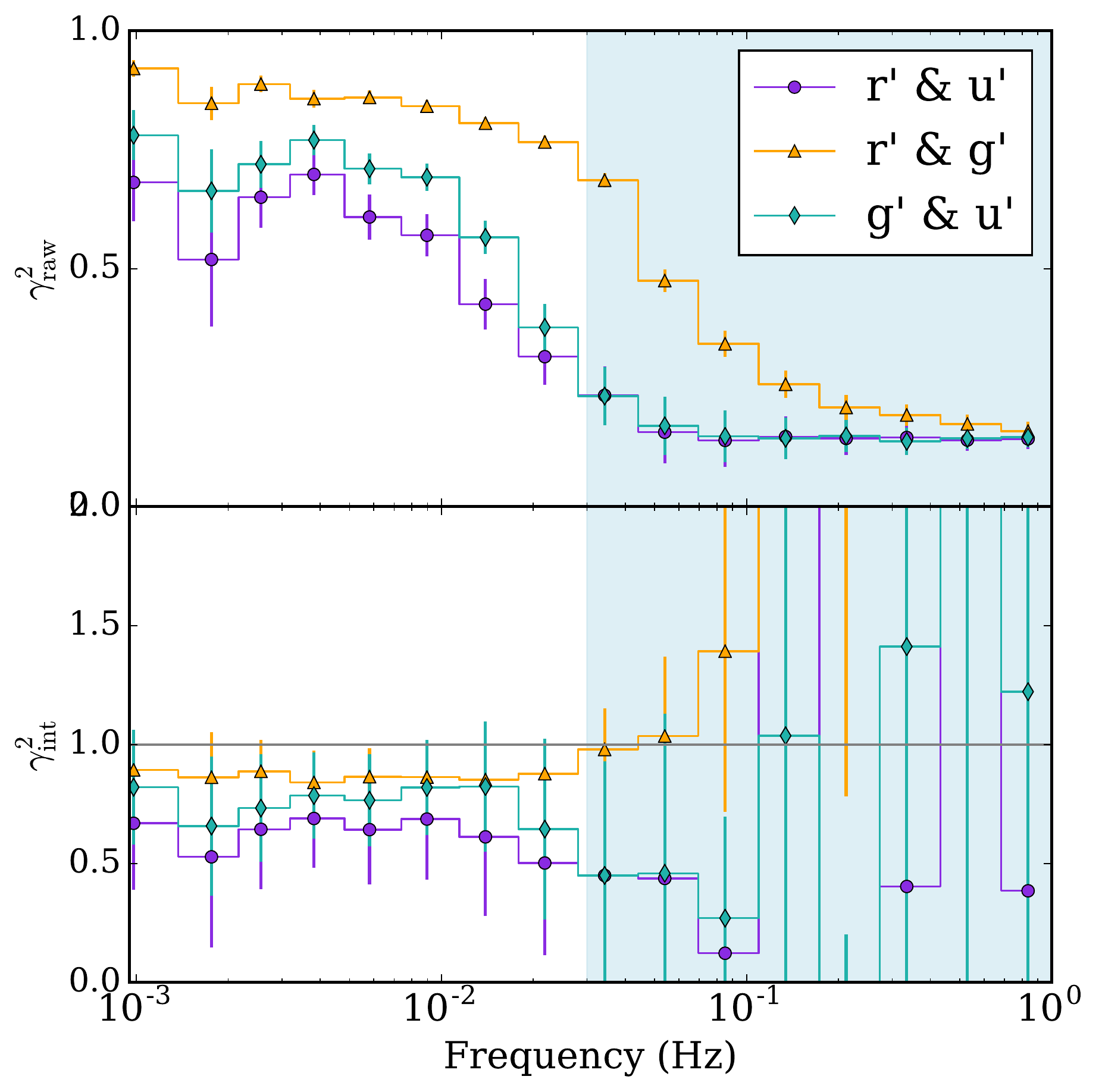}
\caption{Raw and intrinsic coherence for both nights combined, the same symbols are used to denote the different colour combinations. The purple circles denote the $r'$ and $u'$ combination, the green diamonds the $g'$ and $u'$ and the orange triangles the $r'$ and $g'$ combination. The coherence drops at $> 0.01$ Hz. In both figures the shaded region indicates the frequencies that are strongly dominated by Poisson noise in the $u'$ band.}
\label{Fig:combicoher}
\end{figure}

\begin{figure*}
\centering
\includegraphics[width=12.0cm]{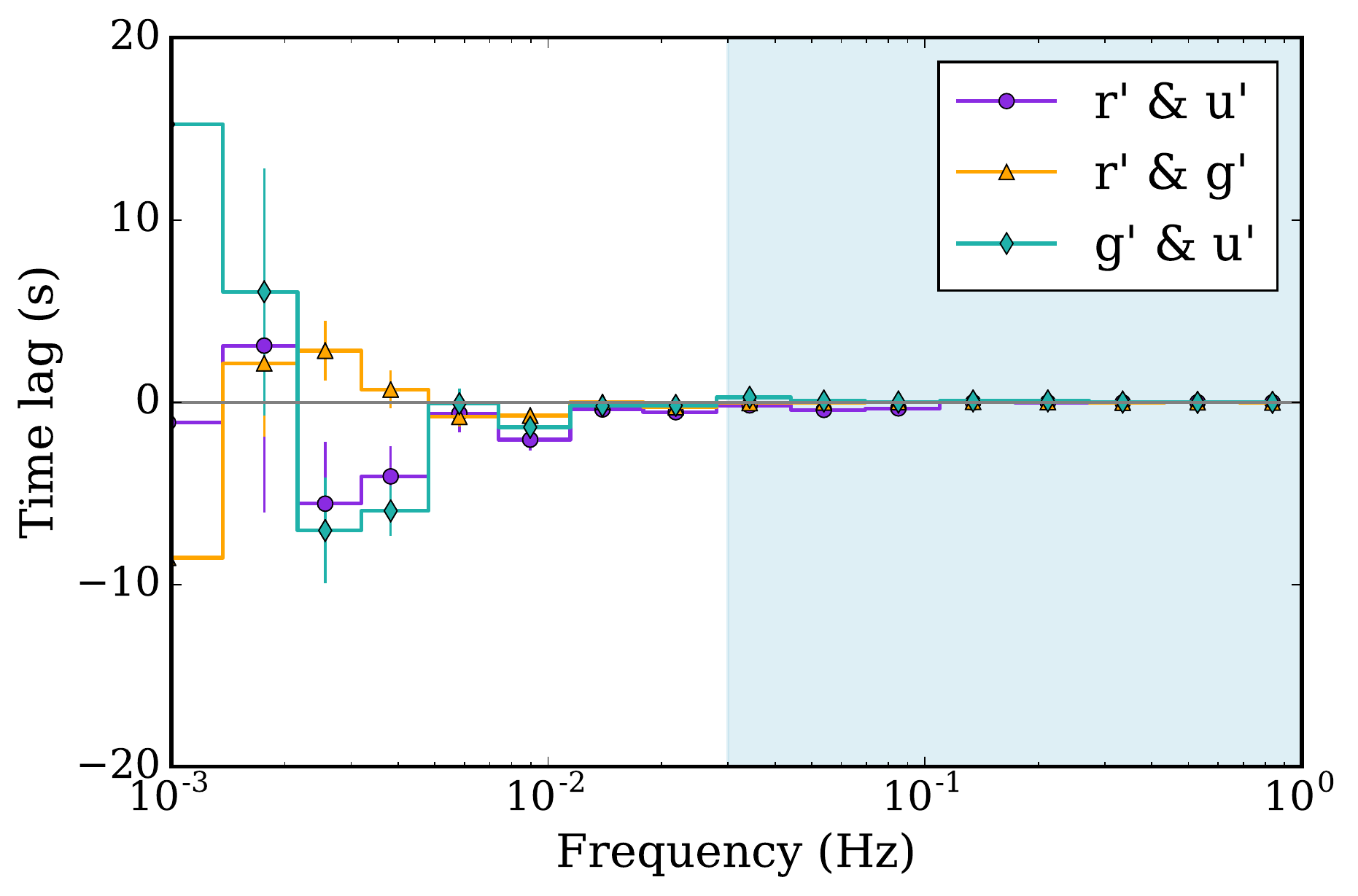}
\caption{Time lags of both nights combined. The purple circles denote the $r'$ and $u'$ combination, the green diamonds the $g'$ and $u'$ and the orange triangles the $r'$ and $g'$ combination. A soft/negative lag is observed between the $r'$ and the $u'$ band as well as the $g'$ and $u'$ bands, the latter is significant at $4\sigma$. The shaded region indicates the frequencies that are strongly dominated by Poisson noise in the $u'$ band.}
\label{Fig:combilags}
\end{figure*}

In Fig. \ref{Fig:lags1} we present the phase/time lags calculated for all the colour combinations. A tentative soft lag of $\sim 5$ s is observed at $\approx4\times10^{-3}$ Hz on both nights in the $r'$ and $u'$, and $g'$ and $u'$ combinations. This indicates that the redder photons are delayed with respect to the bluer photons. In the first night the lag is in one bin and in the second night it appears in several bins at low frequencies. To check whether this is a significant soft time lag we combined the two nights as described in Sect. \ref{sec:Fourier} and we display the result in Fig. \ref{Fig:combilags}. The negative/soft time lags are present in the $g'$ and $u'$, and $r'$ and $u'$ colour combinations at $\approx4\times10^{-3}$ Hz, their amplitude is $-6.0\pm1.4$ s and $-4.1\pm1.7$ s respectively. Hence, the soft lags have a 4$\sigma$ and 2.5$\sigma$ detection in these colour combinations. In Fig. \ref{Fig:combicoher} we show the coherence as a function of frequency (time-scale). The coherence measures the correlation between the time series observed in two different bands at each frequency, thus $1$ is a high correlation and $0$ is no correlation (such as Gaussian noise). The coherence is high at the time-scale where the lag is observed, indicating that the two light curves are well correlated. 

There is no soft lag observed between the $r'$ and the $g'$ band in the results for both nights combined. The positive/hard lag formally observed has a large error bar and is consistent with a zero lag or even mildly negative lags. 
We note that during the first night a consistent soft/negative lag is observed in the $r'$ and $g'$ color combination at the same frequency range where there is a negative lag in the $r'$ and $u'$, and $g'$ and $u'$ combination.
The inconclusive behaviour for the lags between the $r'$ and the $g'$ band will be discussed in Sect. \ref{sec:discussion}. \\

It is interesting to mention the turnover seen from a soft to a hard lag at $2\times10^{-3}$ Hz in the $g'$ and $u'$ colour combination. This behaviour is not seen in the other colour combinations. As mentioned before, the Poisson noise dominates at frequencies above $0.03$ Hz as indicated in Fig. \ref{Fig:psd} and the raw and intrinsic coherence drops, for that reason any apparent phase lag at those frequencies is probably not intrinsic. The frequencies affected by noise are shaded in all the figures. As a test we binned the light curves every 2 seconds and performed the same analysis to check whether the phase lags at higher frequencies disappeared. Whilst the time lag at $\approx4\times10^{-3}$ Hz still holds after binning, we observed that the phase lags at $\approx0.1$ Hz were indeed non meaningful.\\
%

\section{Discussion}
\label{sec:discussion}

We report on the detection of Fourier soft time lags in the DN SS Cygni at $\approx250$ s, with an amplitude of $\sim5$ s ($\approx 20$ cadences) in the $g'$ and $u'$ and $r'$ and $u'$ colour combinations. The soft lag in the $r'$ and $g'$ combination is consistent with a zero lag. As mentioned, the source shows negative lags on the first observing day (see Fig. \ref{Fig:lags1}), and the error bar for the $r'$ and $g'$ colour combination is so large that the findings are consistent with a negative lag, albeit smaller than in the other colour combinations. Part of the reason why the lags in the $r'$ and $g'$ combination are not in agreement with the other colour combinations might be the fact that in the $r'$ band the majority of the flux comes from the companion star, so that the variability coming from the disc is partially masked making difficult the detection of a lag. According to \citet[e.g.][]{MartinezPais1994companion,Harrison2000companion}, the secondary star of spectral type K2 V supplies up to 60 \% of the flux in the $r'$ band ($6500$ \AA), whilst in the $g'$ and $u'$ band the majority of the flux comes from the accretion disc around the white dwarf (see Fig. 10 in \citealt{Harrison2000companion}). A simple constant additive flux will not change the measured lags, only if there is some variability in the added flux, the time lags will change. The K-dwarfs can reach variability amplitudes in the optical bands up to 1-10 milli-mags, or roughly 1 \% \citep[see e.g.][] {Ciardi2011, Basri2013}. Thus, the companion star may add some extra variability predominantly to the $r'$ band and will furthermore reduce the signal-to-noise ratio for the intrinsic variability (due to the added weakly variable flux) and lead to higher uncertainties.

Soft lags have been frequently observed in the X-rays in other compact sources such as XRBs and AGN \citep[e.g.][]{Uttleyreverberation,deMarco2015lags}. In these sources, soft lags have been reported at high frequencies and hard/positive lags at lower frequencies. The hard lags are explained in the context of the propagating fluctuations model, where the harder (bluer) photons arrive later than the softer (redder) photons. The soft lags are thought to represent the light travel time from the corona to the disc, so that the reflected redder photons arrive later than the bluer ones. In the context of an XRB the disc briefly reprocesses the hot photons generated at the corona and re-emits them almost instantly \citep{deMarco2016}. Fig. \ref{Fig:combilags} shows a turnover from soft to hard lags at low frequencies as observed in some AGN and XRBs, but only in the $g'$ and $u'$ colour combination. A tentative positive/hard lag is observed in two consecutive bins, thus it might be significant even if the errors in each bin are larger than those measured for the soft lags. \\

\subsection{Effects on the measured lags}

Certain effects could alter the amplitude of the lag observed at low frequencies. Firstly, the red noise leakage is known to distort the PSD estimates and the phase lags \citep[e.g.][]{vanDerKlis1989,Uttley2002LEAK}. The red noise leakage occurs because of the finite length of the time series. As a result, power from lower frequencies can leak into higher frequencies affecting the variability properties. The effect has been carefully studied by \citet{Alston2013} using simulations of light curves. They show how the leakage can distort the power spectrum and the cross spectrum leading to a bias on the phase lag and the time lag. The amplitude of the soft lag can be reduced up to 50 \%. For a phase spectrum that changes rapidly with frequency the bias on the measurement of the lag will be larger. Whereas, in a smooth phase spectrum the bias should be small. \\

Second, it has been shown in black hole XRBs and AGN that dilution can have a dramatic effect on the amplitude of the observed lags. The measured lags are a weighted contribution of different spectral components, the disc, the driving power-law continuum and the reflection \citep[e.g.][]{Kara2013b,Cackettdilution2013,deMarco2016}. This mixing of variable components in the different colour bands would reduce the measured lag. In recent studies on XRBs and AGN, the amplitude of the intrinsic lag was expected to be 2 times larger than the measured lag. Moreover, according to \citet{Kara2013b}, a low coherence can be an indication of the mixing of components. In the case studied here, the raw coherence is high but it is not exactly 1 at lower frequencies as one would expect. This might suggest that there is a contribution from different components in each optical band: the reprocessed photons, the intrinsic photons from the disc plus the hotter photons generated in the boundary layer or corona. This effect is extremely difficult to account for in CVs because the relative strength of the different variable components is unknown. Generally, in XRBs and AGN the X-ray spectrum of the source can be obtained so that the lags produced by the different emitting components can be simulated \citep[see][for a detailed explanation of this method]{Kara2013b}.\\

The fluxes measured in different optical filters might additionally come from different regions in the disc, and from different optical depths and even levels of excitation. For example, SS Cyg in quiescence shows strong emission lines and the $u'$ bandpass measures the Balmer continuum in contrast to the other bands \citep{spectroSSCyg}. Moreover, it has been shown that in some CVs there is a optical lag between the line emission and the continuum \citep[e.g.][]{Welsh1998}. This illustrates how complex the interpretation of the optical time lags in CVs can be. This option is one possibility that may explain the non-existence of soft lags between the $r'$ and $g'$ bands: If the lags are between the continuum and the emission lines, one would expect to see lags between $u'$ and the other optical bands, but not between the $r'$ and $g'$ bands.
In the following sections we will discuss speculative physical scenarios to explain the negative/soft lags reported here in SS Cygni and investigate the relevant time-scales that might play a role. We will further discuss how the soft lags presented in this work relate to the recent detections of lags in other CVs.

\subsection{Geometrical interpretation}
\begin{figure}
\centering
\includegraphics[width=8.0cm]{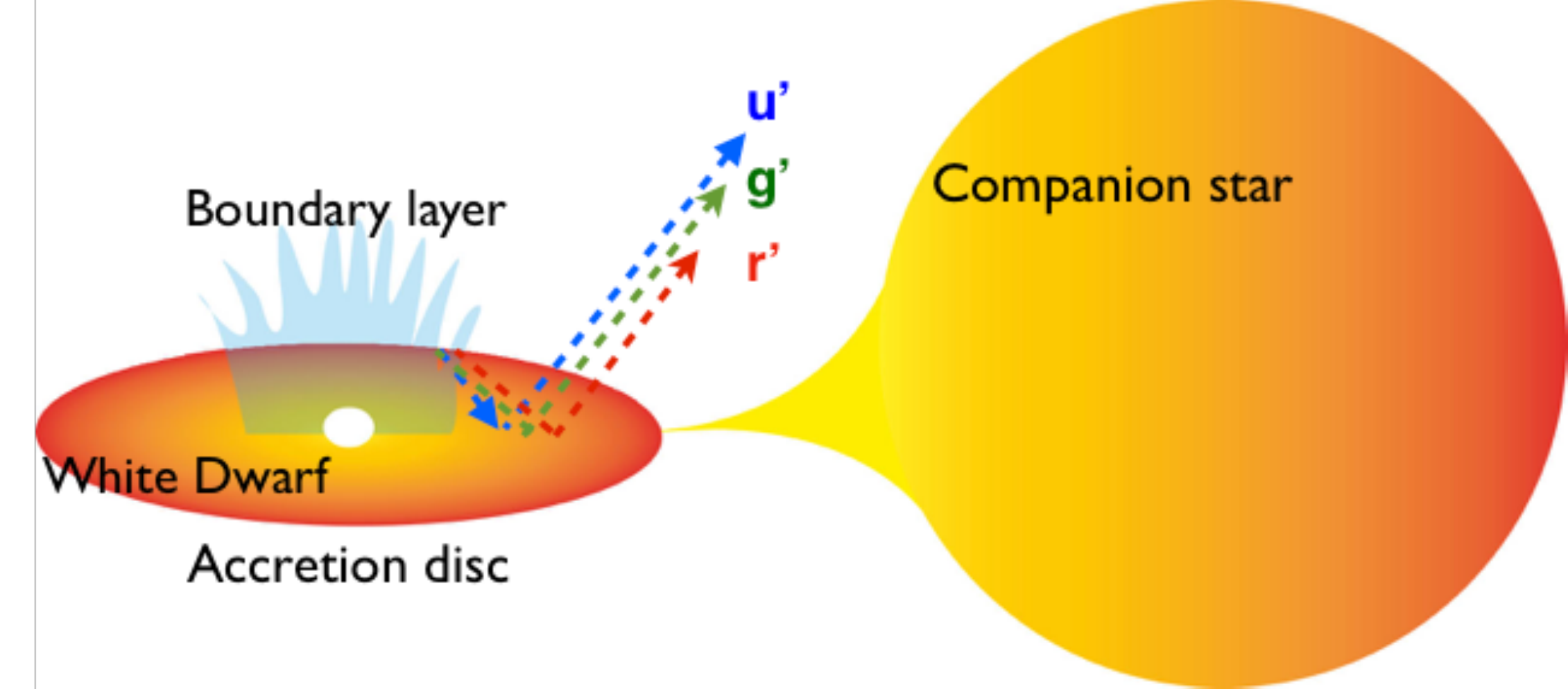}
\caption{Cartoon showing the reprocessing of high-energy photons generated in the boundary layer in the accretion disc. The bluer photons are reprocessed closer to the inner edge of the disc and re-emitted faster so that the redder photons arrive with a delay of seconds.}
\label{Fig:piclags}
\end{figure}

The time-scale of 4.2 min where the lag appears can be associated with the dynamical time-scale of the disc at a radius of $\sim10\,R_{WD}$. For a system like SS Cyg with a large orbital period of $6.6$ h, the disc is expected to be quite large; in particular the circularisation radius where the material starts to follow Keplerian orbits is expected to be at $30\,R_{WD}$. The reason why the lag appears on that time-scale is still puzzling, but we can speculate that this might be a region where the disc becomes optically thin. \\

To explain the observed amplitude of $4-6$ s, we could potentially associate it with the light travel time and instant reprocessing of photons like in XRBs. However, the reflecting region would lie between $2.5-3\,\rm R_{\odot}$ and this region is outside the accretion disc, since the semi-major axis of the orbit of SS Cyg is $1.95\, R_{\odot}$,$\sim4.5$ light seconds (calculated using the binary parameters derived in \citealt{Bitner2007}). Similar conclusions were drawn by \citet{Scaringi2013}; the soft lags observed in LU Cam could not be explained by the light travel time because the emitting region would lie outside of the disc. In the case of CVs, the accretion disc is much cooler than discs in XRBs, and this can be translated into a much longer reprocessing time for the CVs. We can propose two different toy models to explain the reprocessing of photons.\\

\subsubsection{Thermal time-scale}

If the reprocessing occurs on a thermal time-scale, as proposed by \citet{Scaringi2013}, we need sufficient irradiation to alter the local thermal equilibrium on the surface of the disc. The thermal time-scale is:
 
\begin{equation}
\rm t_{\rm th} = \frac{t_{\phi}}{\alpha} = \frac{1}{\alpha}\left(\frac{G\,M_{\rm WD}}{R^{3}}\right)^{-1/2},
\end{equation}

\noindent where $\alpha$ is the viscosity parameter and $R$ is the disc radius \citep{Shakura1973}. If the conditions are such that the energy balance of a specific area in the disc changes due to irradiation of high energy photons, the time-scale for re-adjustment to thermal equilibrium can be associated with the amplitude of the lag. If we associate the observed 5 s lag with the thermal time-scale and use the estimated mass $\rm M_{\rm WD}=0.81\,\rm M_{\odot}$ derived in \citet{Bitner2007}, this would place the reprocessing region between $0.01-0.02\, R_{\odot}$ for $0.3<\alpha<0.7$. This region is at the inner-most edge of the accretion disc, very close to the WD surface. Hence, for viscosity parameters $\alpha > 0.3$ the soft lags of SS Cyg could potentially be explained by the thermal time-scale, so that the higher energy photons are reprocessed very close to the WD and re-emitted after $\sim 5$ s (for a toy model of this process see Fig. \ref{Fig:piclags}). However, values of $\alpha> 0.3$ are higher than those predicted for a DN in quiescence, such as SS Cygni at the time of the observations. Conversely, in quiescence $\alpha$ is expected to be $\approx0.02$, and it can increase by up to one order of magnitude during outburst \citep{Frank2002}. Lower values of the viscosity parameter would then place the reprocessing region too close to the white dwarf, so that the soft lags could not be associated with a thermal time-scale.  
Nevertheless, there have been claims of a higher viscosity parameter obtained using the flickering mapping technique on V2051 Oph of $\alpha_{\rm quiesc} = 0.1-0.2$ \citep[e.g.][]{Baptista2004}. Furthermore, for the CV HT Cas a value of $\alpha_{\rm quiesc} = 0.3-0.5$ was reported \citep[e.g.][]{Baptista2011,Scaringi2014}. Furthermore, if the effects of the red noise leakage or the dilution are indeed important, the amplitude of the real lag should be slightly larger. Hence, thermal reprocessing of photons in the disc can be a tentative explanation for the observed soft lags in SS Cyg, but only under conditions of high irradiation and high viscosity.\\

\subsubsection{Recombination time-scale}

Another relevant time-scale that can play an important role in the reprocessing of photons is the recombination time-scale. The recombination time-scale is: $t_{rec}\sim(n_{e}\alpha_{rec})^{-1}$ seconds \citep{obrien2002}, where $n_{e}$ is the electron density, typically $n_{e}\sim 10^{13}\,cm^{-3}$ at the mid-plane of the accretion disc in CVs \citep[e.g.][]{Warner2003}, and $\alpha_{rec}$ is the recombination coefficient. According to \citet{Hummer1963} the recombination coefficient can be roughly estimated using this analytical expression:

\begin{equation}
\alpha_{rec}= 1.627\times10^{-13}t_{e}^{-1/2}(1-1.657log_{10}t_{e}+0.584t_{e}^{1/3}),
\end{equation}

\noindent where $t_{e}=10^{-4}T_{e}$, and $T_{e}$ is the temperature of the disc. For a temperature of $T_{e}\sim10^{4}$ K $\alpha_{rec}=2.5\times10^{-13}\,cm^{3}s^{-1}$. This yields a recombination time-scale of $0.4$ s, smaller than the amplitude of the lag reported here. However, if the recombination occurs in the upper layers of the disc where the density is lower, the recombination time-scale can be adequate to explain the amplitude of the lag seen. At a vertical distance of 2H, where H is the scale height of the disc, the density is 10 times lower, this will mean a 4 s lag. This time-scale has been considered to explain the time lag observed in XRBs between the optical/UV and the X-ray emission \citep[e.g.][]{Hynes1998}. The high energy photons irradiate the disc and are reprocessed and re-emitted in the optical wavelengths in the outer parts of the disc.

\subsection{Comparison with previous detections of soft time lags in CVs}

Soft/negative lags in SS Cyg were reported by \citet{Bruch2015lags} with a similar amplitude to those observed here. The authors used cross-correlation functions as they consider it more sensitive to detect lags at low frequencies. For comparison we have computed the cross-correlations and found similar soft lags. In this work we used advanced Fourier analysis because it enables us to measure the time lags produced at different time-scales. Thus, in the frequency domain we could detect hard and soft lags as observed in XRBs and AGN. Similar soft lags were also reported for the CVs V603 Aql, TT Ari and RS Oph by \citet{Bruch2015lags}. In addition, Fourier soft lags have been reported in \citet{Scaringi2013}. Thus, with the detection of the soft time lags in SS Cyg there are now three CVs showing Fourier soft time lags, and including the lags detected using the CCFs there are six CVs. The amplitude of the lags found here are similar to the soft lags of MV Lyr and LU Cam, with time delays of $\sim3$ seconds and $\sim10$ seconds respectively. Moreover, the SS Cygni lags are present at a similar time-scale, $\approx250$ s to the soft lags in LU Cam and MV Lyr. \

Given the similarities observed between the soft lags of these three CVs, we believe that the same physical process is creating the negative lags. In \citet{Scaringi2013} the lags were explained by thermal reprocessing of photons for high values of viscosity $\alpha=0.7$. In addition, they discuss a different origin also associated with the thermal time-scale but from reverse shocks within the accretion disc. The difference seen in amplitude of the lags for these CVs cannot be explained by the different estimated masses of the white dwarfs, but accurate masses are difficult to obtain. The difference cannot be associated to other parameters of the system such as the orbital period. The orbital period of MV Lyr is $3.19$ h \citep{SkillmanperiodMVLyr}, for LU Cam $3.6$ h \citep{SheetsperiodLUCam} and for SS Cyg $6.6$ h. Even though the latter is much larger than the other orbital periods, the amplitude of the lag in SS Cyg is slightly shorter than the one in LU Cam. \\

Further investigation of other CVs is required to confirm the presence of soft lags in other sources. With a larger sample it would be possible to search for a correlation of the lags with the system parameters, helping us to understand the physical processes occurring in the accretion discs around white dwarfs. It is equally important to explore whether the lags change sign to positive/hard lags at lower frequencies as observed in XRBs and AGN. For this purpose it would be necessary to monitor the source for a longer time to reach the lower frequencies. Therefore, in order to detect hard lags in CVs, optical instruments in space with the ability to observe in different optical filters simultaneously are required. With PLATO it would be possible to observe in a red and a blue filters simultaneously\footnote{http://sci.esa.int/plato/59252-plato-definition-study-report-red-book/}. This goal is difficult to achieve using ground-based telescopes, as we cannot observe these sources continuously for many hours. It could be possible to use Las Cumbres Observatory as it will enable us to observe continuously for longer time using the telescopes located in different countries in the world.  \\

\section{Summary}
\label{sec:summary}

In this work we have performed observations of SS Cygni with ULTRACAM, enabling us to observe it in three filters simultaneously at very high temporal resolution. We have analysed the light curves in the Fourier domain and derived the power spectra, coherence and phase and time lags. We find significant lags of $\sim 5$ s in the $g'$ and $u'$ and $r'$ and $u'$ colour combinations at a time-scale of $\approx250$ s ($\approx4\times10^{-3}$ Hz). This confirms the existence of soft lags in SS Cyg suggested by \citet{Bruch2015lags}. The amplitude of the lags as well as the time-scale where they were observed are consistent with the lags observed in other CVs such as LU Cam and MV Lyr. We propose that the nature of the soft lags could be explained by the reprocessing of higher energy photons coming from the boundary layer in the accretion disc. We considered two different physical processes that could produce the observed amplitude of the lag, namely a thermal re-adjustment of the disc on a thermal time-scale and the recombination of the ionised gas in the disc. The first possibility requires sufficient irradiation and also a high viscosity parameter $\alpha>0.3$. In this case the reprocessing region would be at a radius of $0.01-0.02\, R_{\odot}$. Such high values of $\alpha$ are not expected for a DN in quiescence. The recombination time-scale on the surface of the disc where the density is lower can also explain the amplitude of the lag reported here. Furthermore, apart from these geometrical scenarios to explain the soft lags, it is important to remark that the difference between the flux in the $u'$ and the $g'$ band is not only bluer versus redder, and then hotter versus colder, it is also probing different optical depths and excitation. The study of Fourier time lags can give insights into the physical processes in accretion discs and help us to constrain important parameters of these accreting systems. 

\section*{Acknowledgements}

This work is partly financed by the Netherlands Organisation for Scientific Research (NWO), through the VIDI research program Nr. 639.042.218. VSD and ULTRACAM are supported by the STFC. We thank the anonymous referee for the comments and suggestions that helped to greatly improve this article.



\bibliographystyle{mnras}
\bibliography{sscyg} 

%


\bsp	
\label{lastpage}
\end{document}